\documentclass[fleqn,10pt]{wlscirep}
\usepackage[utf8]{inputenc}
\usepackage[T1]{fontenc}
\title{Purcell-Enhanced Single Photons at Telecom Wavelengths from a Quantum Dot in a Photonic Crystal Cavity}
\author[1,*]{Catherine L. Phillips}
\author[1]{Alistair J. Brash}
\author[2]{Max Godsland}
\author[1]{Nicholas J. Martin}
\author[1]{Andrew Foster}
\author[1]{Anna Tomlinson}
\author[1]{René Dost}
\author[2]{Nasser Babazadeh}
\author[2]{Elisa M. Sala}
\author[1]{Luke Wilson}
\author[2]{Jon Heffernan}
\author[1]{Maurice S. Skolnick }
\author[1]{A. Mark Fox}

\affil[1]{Department of Physics and Astronomy, University of Sheffield, UK}
\affil[2]{EPSRC National Epitaxy Facility, Department of Electronic and Electrical Engineering, University of Sheffield, UK}

\affil[*]{c.l.phillips@sheffield.ac.uk}

\begin{abstract}
 Quantum dots are promising candidates for telecom single photon sources due to their tunable emission across the different low-loss telecommunications bands, making them compatible with existing fiber networks. Their suitability for integration into photonic structures allows for enhanced brightness through the Purcell effect, supporting efficient quantum communication technologies. Our work focuses on InAs/InP QDs created via droplet epitaxy MOVPE to operate within the telecoms C-band. We observe a short radiative lifetime of 340 ps, arising from a Purcell factor of 5, owing to integration of the QD within a low-mode-volume photonic crystal cavity. Through in-situ control of the sample temperature, we show both temperature tuning of the QD's emission wavelength and a preserved single photon emission purity at temperatures up to 25K. These findings suggest the viability of QD-based, cryogen-free C-band single photon sources, supporting applicability in quantum communication technologies.
\end{abstract}
\begin{document}

\flushbottom
\maketitle
%
%
\thispagestyle{empty}


\section*{Introduction}

Many quantum communication and quantum information processing applications require a source of indistinguishable single photons \cite{munro2012quantum,azuma2015all,PhysRevLett.86.5188,walther2005experimental,rudolph2017optimistic}. There are many possible platforms for creating single photon sources (SPS), including colour centres in materials such as diamond \cite{benedikter2017cavity} or silicon \cite{redjem2023all}, 2D material emitters \cite{doi:10.1021/acs.nanolett.2c03151} and trapped ions \cite{Barros_2009}. One of the most promising SPS candidates is the semiconductor quantum dot (QD)\cite{Heindel2023}, where a large dipole moment and relatively weak phonon coupling provide attractive optical properties for the realisation of bright single photon sources. Furthermore, the potential to integrate QDs into semiconductor nanostructures offers additional desirable functionalities; for example, photonic crystal cavities (PhCCs) to enhance emission rates and waveguides for on-chip photon routing and to exploit chiral\cite{Siampour2023} and topological\cite{JalaliMehrabad2020} properties. From an SPS perspective, PhCCs can significantly enhance the emission properties of the source via the Purcell effect, in the form of higher brightness in the weak-coupling regime and reduced sensitivity to environmental noise.

Notably, semiconductor QDs may be engineered to emit over a large spectral emission window, ranging from the visible to the technologically relevant telecommunications C-band (1530-1565 nm). While the highest performance QD SPS currently operate in the Near-Infrared (NIR) (800-1000nm), the potential for operation in the low-loss C-band spectral window for standard silica fiber provides strong motivation for their continued development. Such a source would be compatible with existing fiber networks, with the potential for long distance fiber-based quantum communication \cite{Vajner2022} as well as low-loss connectivity between individual nodes of the future quantum internet. 

Alongside wavelength compatibility, integration into the existing network  requires fiber coupling of sources. 
Work in similar semiconductor membrane platforms has shown the possibility of compact, high efficiency devices utilising fibre-coupled gratings or tapered waveguides\cite{coupler_review}.  The clear path to high-efficiency fiber coupling is a motivation behind our PhCC approach\cite{lodal_sps,lodahl_grating,Far_field_1550}.



To date, several approaches have been investigated to realise QD SPSs which emit directly within the C-band. One such approach involves growing GaAs QDs on an InGaAs metamorphic buffer layer \cite{Ledentsov2007,Semenova2008,Sittig2022,Paul2017,Wroski2021} in order to reduce the lattice mismatch between the QDs and the growth substrate, allowing larger QDs to form and thereby shifting the emission from $\sim$900 nm to the telecom C-band. The second method involves growing InAs QDs on InP\cite{Benyoucef2013,Mller2018,Skiba2017} either using the Stranski-Krastanow (SK) self-assembly approach or droplet epitaxy (DE). In this work we focus on the latter approach, using DE in a metal-organic vapour phase epitaxy (MOVPE) reactor. 

DE was initially developed and widely studied in Molecular Beam Epitaxy (MBE) \cite{KOGUCHI1991688,Gurioli2019}, particularly using the InAs/GaAs system, but has shown significant promise as an MOVPE technique. Recently, the first quantum-light emitting diode (QLED) based on telecom C-band InAs/InP DE QDs grown by MOVPE was demonstrated \cite{Mller2018}. Such QDs were reported to have a lower fine-structure splitting (FSS) and longer coherence times than their counterparts grown by the SK method \cite{Skiba2017,Anderson2021}. They are thus promising candidates for highly indistinguishable single and entangled photon sources. Following the demonstration of the first telecom C-band QLED, reports discussed the growth mechanism, detailed morphology and optical characterization of InAs/InP DE QDs in MOVPE \cite{Sala2020,Sala2021,Sala2021IOP,Gajjela2022,Sala2023}.These showed the possibility to fabricate and engineer defect-free, symmetric, and pure InAs QDs emitting in the telecom C-band using the industrially compatible MOVPE technique.

As mentioned previously, a particular advantage of semiconductor QDs for SPSs is that they can be straightforwardly integrated into photonic structures such as cavities. In the QD-cavity weak-coupling regime, the Purcell effect can be harnessed to decrease the radiative lifetime of the emitter, allowing for higher repetition rate driving and therefore increasing the SPS brightness. 
A short lifetime is desirable for high clock rate quantum key distribution (QKD) and quantum communications, enabling increased secure key rates over long distances \cite{Waks2002,Morrison2023}.

The Purcell factor ($F_P$) of a point emitter in a cavity is dependent on the spectral, spatial and polarisation overlap of the cavity mode and the relevant QD transition as well as the cavity quality factor (Q), and is given by\cite{Liu2018}:
\begin{equation}
    F_P = \frac{\tau_{bulk}}{\tau_{cav}} = \frac{(3Q)}{(4\pi^2 V)} \frac{\omega_c^2}{4Q^2 (\omega-\omega_c)^2+\omega_c^2} \epsilon^2,
    \label{eqn:Purcell}
\end{equation}
where $\tau_{bulk}$ is the radiative lifetime of the QD in the absence of a cavity and $\tau_{cav}$ is the radiative lifetime of a QD in a cavity, Q is the quality factor of the cavity, V is the cavity mode volume, $\omega$ and $\omega_{c}$ are the QD transition and cavity mode frequencies respectively and $\epsilon$ quantifies the spatial overlap between the QD dipole and the cavity mode.

Here, we report on a QD embedded within an L3 PhCC emitting very close to the telecoms C-band. We show that the QD can be driven using quasi-resonant excitation via the LA-phonon sideband, and demonstrate temperature tuning of the Purcell enhancement, with a short radiative lifetime of $\sim$340 ps measured on resonance with the cavity (corresponding to $F_{P} \approx$ 5). We also show that the purity of the single photon emission is unchanged up to at least 25K, which is promising for future operation of a C-band SPSs under cryogen-free cooling.

\section*{Results}

Our PhCC samples were installed in a liquid helium bath cryostat at 4.2K and investigated using a combination of continuous wave (CW) and pulsed laser excitation both above-band and quasi-resonantly through the LA-phonon sideband. More details of the excitation schemes used can be found in the "Methods".

\begin{figure}[ht]
\centering
\includegraphics[width=\linewidth]{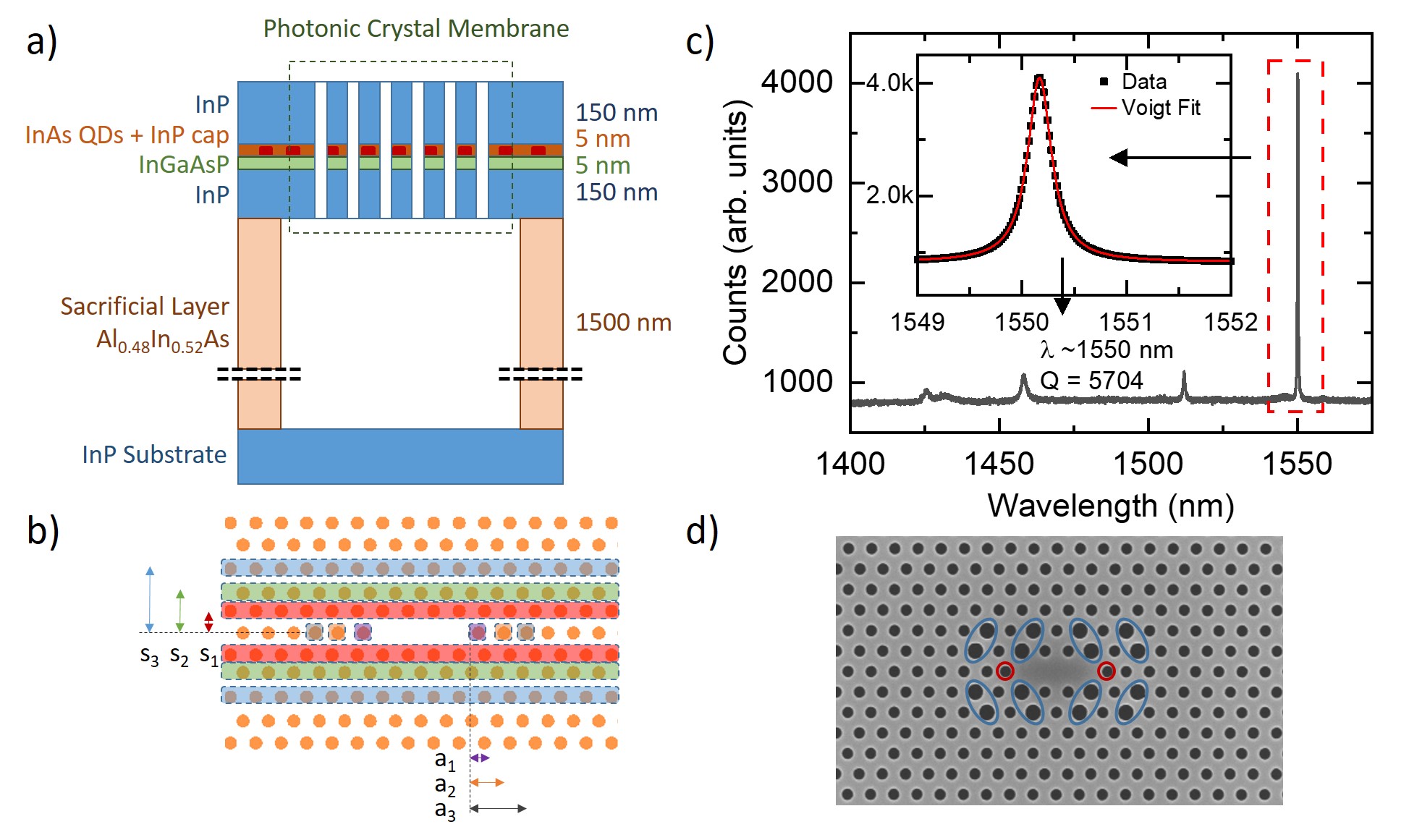}
\caption{a) Schematic cross-section of the QD semiconductor wafer. InAs QDs are grown via droplet epitaxy in MOVPE within a 310 nm thick InP membrane on a 5 nm InGaAsP interlayer. Removal of the AlInAs sacrificial layer during fabrication results in a suspended membrane photonic crystal structure. The depth of the sacrificial layer has been chosen using FDTD simulation to maximise collection from above the wafer in the telecom C-band, due to constructive interference caused by reflections from the InP substrate. b) Schematic of an L3 cavity optimised for a high Q factor. The highlighted areas represent the rows and holes moved using guided mode expansion to maximise the simulated Q factor of the cavity. c) High power non-resonant measurement of L3 cavity modes in a PhCC designed to have a high Q factor. The fundamental mode here is at $\sim$1550 nm with a Q of 5704. The four modes shown match the mode spacing simulated using GME. Inset: Magnification of the fundamental cavity mode measurement (black squares) showing the Voigt fit (red line) from which the Q is calculated. d) Scanning electron microscope image of a fabricated L3 cavity that has been optimised for far-field collection. The blue and red circles outline the sets of holes where the radii and lattice constant has been modified to optimise the Gaussian far-field. }
\label{fig:fig_1}
\end{figure}

\subsection*{Photonic crystal cavity devices}
 
Figure (\ref{fig:fig_1}a) shows the membrane structure used in our PhCC devices, in which the suspended InP membrane has a thickness of 310nm, with MOVPE grown DE QDs grown in the centre on a thin 5nm InGaAsP interlayer\cite{Sala2023}. Below the membrane there is a 1500 nm thick AlInAs sacrificial layer, which is etched away during device fabrication to leave a suspended membrane PhCC. The thickness of the sacrificial layer was optimised using finite-difference time-domain (FDTD) simulation to maximise collection from above the wafer in the telecom C-band, due to constructive interference caused by reflections from the InP substrate. 

\subsubsection*{High-Q L3 cavity using GME based inverse design}

Inverse design has been shown to be a powerful tool for the design optimisation of nano-photonic structures \cite{Molesky2018}. Here, we used the inverse design capabilities of the Legume guided mode expansion (GME) software package \cite{Minkov2020} to maximise the Q factor of an L3 PhCC. We began with a lattice of period a = 425nm and hole radius r = 115nm and focused first on tuning the position of those holes closest to the cavity, with the resulting modifications highlighted in Figure (\ref{fig:fig_1}b). The rows labelled s1, s2 and s3 were optimally shifted by 1a, 0.94a and 0.97a, respectively, with the shifts being mirrored with respect to the cavity axis. The first three holes at the ends of the cavity were also displaced outwards from the cavity by a1 = 1.13a, a2 = 1.09a and a3=1.04a, respectively. 

\begin{figure}[ht]
\centering
\includegraphics[width=\linewidth]{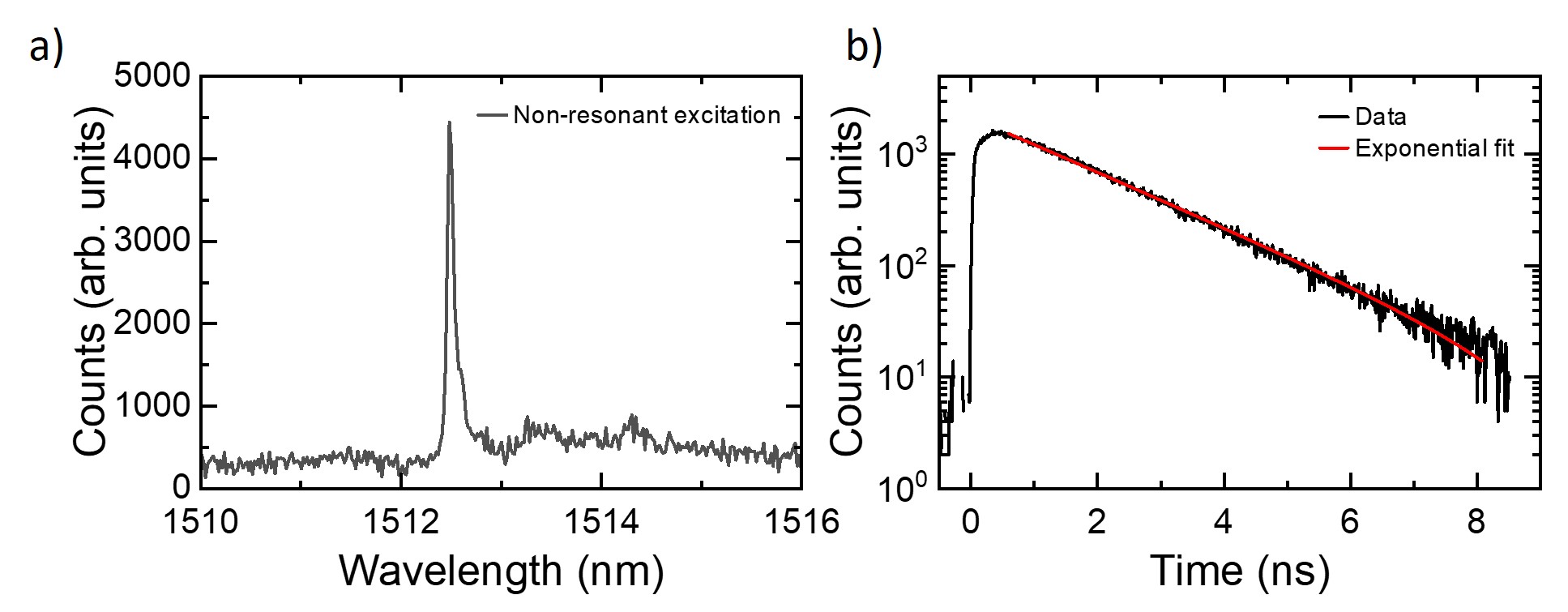}
\caption{a) Non-resonant exciation PL spectrum of an InAs QD in an InP membrane with a wavelength of 1512.48 nm b) Time-resolved non-resonant lifetime measurement of the QD shown in (a), fitting gives a radiative lifetime of 1.78$\pm$0.01 ns. }
\label{fig:fig_2}
\end{figure}

The passive optical properties of the fabricated device were investigated with the sample mounted in a liquid helium bath cryostat and interrogated using high-power non-resonant excitation ($\lambda$ = 850 nm). This populated the four lowest order modes of the cavity through broadband emission generated by the QD ensemble. A representative mode spectrum is shown in Figure (\ref{fig:fig_1}c), in which the cavity has a fundamental mode wavelength of $\sim$1550 nm and Q factor of 5,700. GME was found to reproduce the splitting between the first three cavity modes to within 5$\%$, 
showing that the GME approach can be used to greatly increase the speed of large parameter sweeps compared with the more commonly used FDTD. The highest Q factor measured across the sample was 10,000 (at a fundamental mode wavelength of 1525nm). The measured values are therefore notably smaller than the simulated Q factors of $\sim$ 100,000. This is possibly due to the relatively thick membrane used in this work (310 nm) which will increase the sensitivity of the Q factor to imperfect sidewall verticality or accompanying roughness.

\subsubsection*{Improving far-field collection}
In ref \cite{Portalupi2010} it is shown that the extraction efficiency from above an L3 PhCC can be improved by modifying both the radius and position of holes closest to the cavity (albeit at the expense of a reduction in the Q factor). As a step in this direction, a second cavity design was investigated.

Figure (\ref{fig:fig_1}d) shows a scanning electron microscope (SEM) image of the resulting L3 PhCC cavity design, in which we modified the Portalupi et al.\cite{Portalupi2010} design using GME and FDTD for an InP wafer membrane thickness of 310 nm and a sacrificial layer thickness of 1500 nm. The blue circles mark the holes where the radius has been increased by 20-40nm to improve the far field Gaussian properties of L3 cavity and the red holes have been modified by a shift of 0.2a away from the cavity \cite{Far_field_1550} to increase the quality factor\cite{Portalupi2010}.

\subsubsection*{Purcell enhanced QD emission}

Next, we investigate the potential of the far-field-optimised device as a single photon source.
We first obtained a reference radiative lifetime ($\tau_{bulk}$) using QDs in a bulk wafer. These measurements were undertaken using non-resonant pulsed excitation at 850nm. The QD emission was filtered through a spectrometer with a filter bandwidth of $\sim$60 µeV. An example bulk QD spectrum is show in Figure (\ref{fig:fig_2}a) for a QD with emission wavelength of 1512.48 nm and linewidth of 38 µeV (note that linewidths down to $\sim$20 $\mu$eV, the resolution of our spectrometer, were measured on this sample). Figure (\ref{fig:fig_2}b) shows the measured time-resolved non-resonant lifetime measurement for the QD shown in Fig. (\ref{fig:fig_2}a). A single exponential fit is used to extract $\tau_{bulk}$ =1.78 ± 0.01 ns. Multiple single QDs were measured with lifetimes ranging from 1.6-1.8 ns. This agrees with previously published radiative lifetime measurements for QDs emitting in the telecom C-band \cite{Nawrath2021,Anderson2021,Paul2017,Wroski2021,Tomimoto2007}.

\begin{figure}[ht]
\centering
\includegraphics[width=0.875\linewidth]{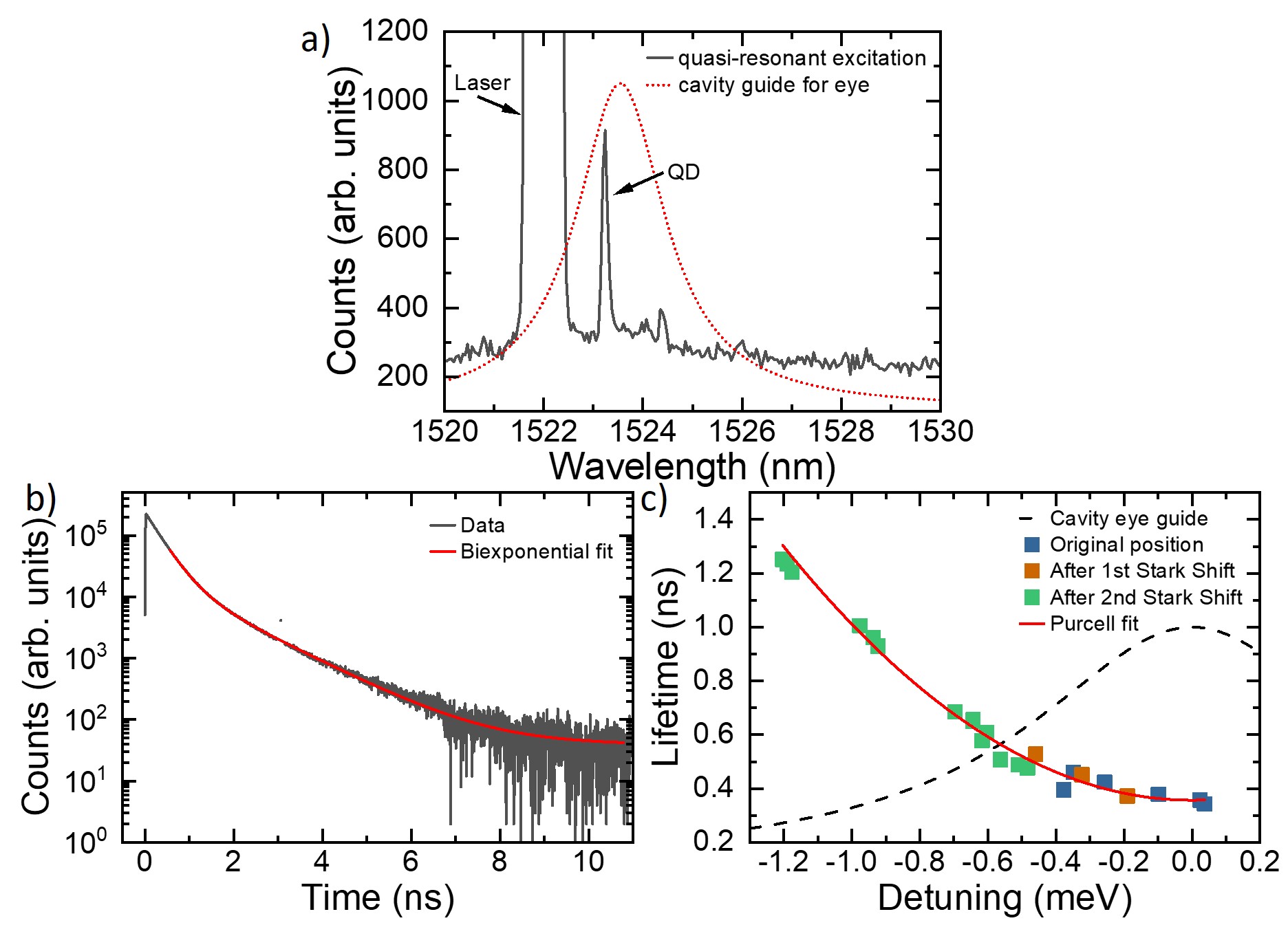}
\caption{a) PL spectrum of a QD in an L3 cavity where the QD is excited via a pulsed OPO through the phonon sideband of the QD at 1522 nm (labelled "laser"). The red line shows the spectral position and shape of the cavity mode. At 4K, shown here, the QD is blue detuned from the centre of the cavity at 1523.2 nm. b) Time-resolved radiative decay (grey line) of a QD measured under phonon sideband pulsed excitation at 4K, fitted with a biexponential fit (red line). The lifetime measurement shows three components. A fast detector-response limited component caused by non-radiative emission (which is not fitted using the biexponential). The two components of the biexponential describe $\tau_{cav}$ of the QD, dependent on the cavity detuning, and a time component that matches literature values for $\tau_{bulk}$ and is not cavity detuning dependent. c) The $\tau_{cav}$ component of the QD radiative lifetime measurement is plotted against QD detuning from the centre of the cavity. The blue, orange and green points represent the temperature and excitation power dependent measurements performed for the three times the QD wavelength DC Stark shifted at 4K. The cavity mode wavelength dependence is shown as a guide for the eye (black dashed line). The data is fitted using Equation \ref{eqn:Purcell} (red line).}
\label{fig:fig_3}
\end{figure}

We then identified a QD transition which was resonant with the fundamental mode of an L3 cavity at 1523.2 nm, using quasi-resonant LA phonon sideband excitation\cite{PhysRevLett.114.137401} with a laser-QD detuning of 0.67 meV. The resulting PL spectrum is shown in Figure (\ref{fig:fig_3}a). We note that this particular transition was not observed under non-resonant, above band excitation at 850 nm, 
illustrating that the presence/absence of additional itinerant photoexcited carriers appears to modify the QD charge state. A weak background signal can also be seen in Fig (\ref{fig:fig_3}a) as well as the QD emission. Next, the radiative lifetime of the QD was determined under pulsed phonon sideband excitation, using a pulse duration of $\sim$20 ps. The excitation laser was rejected using a cross-polarisation scheme, with the collected emission subsequently filtered using a grating filter with a bandwidth of 0.1 nm, and detected using a superconducting nanowire single photon detector with a jitter time of 25 ps. 
The resulting decay curve is shown in Figure (\ref{fig:fig_3}b) and comprises three components. 
The first is a fast component that is instrument response limited; which we attribute to the broad background observed in the QD spectra. The second component of the decay curve is attributed to the QD decay. When the QD is resonant with the cavity we find $\tau_{cav}$ = 0.343$\pm$ 0.001 ns. Comparing this to $\tau_{bulk}$ using using Equation \ref{eqn:Purcell} gives a $F_P$ of 5.  
A final slow component of the decay has a time constant of 1.2 ns, which we attribute to other QDs more weakly coupled to the cavity but lying spectrally within the bandwidth of our filter.

\subsubsection*{Detuning dependence of the Purcell factor}
To investigate the Purcell factor further, we measured $\tau_{cav}$ whilst tuning the QD transition relative to the cavity mode. This tuning was achieved using both the sample temperature and the power of the excitation laser (see Methods). We also observed that the QD transition energy periodically shifted over time to longer wavelengths, allowing the effect of the cavity on the radiative lifetime to be mapped over a larger range of QD-cavity detunings. We hypothesise that the slow drift is also due to a DC Stark shift caused by charge instability near the QD. The absence of a diode structure with electrical control means the current sample has no mechanism to apply an external electric field to stabilise the charge environment.

Using the combination of tuning approaches, we mapped the Purcell factor from on-resonance up to a detuning of -1.2 meV. Figure (\ref{fig:fig_3}c) shows $\tau_{cav}$ as a function of QD-cavity detuning. The blue, orange and green points represent the temperature and excitation-power-dependent lifetime measurements performed at three different reference wavelengths of the QD transition (i.e. under low power excitation at 4.2K). The black dashed line is a guide for the eye and shows the spectral shape of the cavity mode.

It is evident that as the QD transition is tuned away from resonance with the cavity mode, the radiative lifetime increases towards the value measured for QDs in bulk. Fitting the detuning dependent lifetimes using the weak cavity-emitter coupling model of Equation \ref{eqn:Purcell}, where the cavity and QD parameters $\omega_c$, $\omega$ are known, $\tau_{bulk}$ = 1.7 and where the mode volume of an L3 PhCC is on the order of 0.8 ($\lambda/n)^3$ \cite{Rickert2020,Hennessy2007,Chalcraft2007,Akahane2003}, the spatial overlap between the QD and the cavity mode ($\epsilon$) of $\sim$ 0.3 can be derived. The excellent agreement between this fit (red line in Figure \ref{fig:fig_3}(c)) and the experimental data provides further evidence that the measured lifetimes originate from the Purcell enhancement of the cavity. By contrast, the longer-timescale component of the lifetime measurements did not show a dependence on the QD-cavity detuning, supporting the suggestion that this component is a contribution from QDs with very weak spatial overlap with the cavity mode.

\subsubsection*{Hanbury-Brown and Twiss second-order correlation measurements}
\begin{figure}[ht]
\centering
\includegraphics[width=\linewidth]{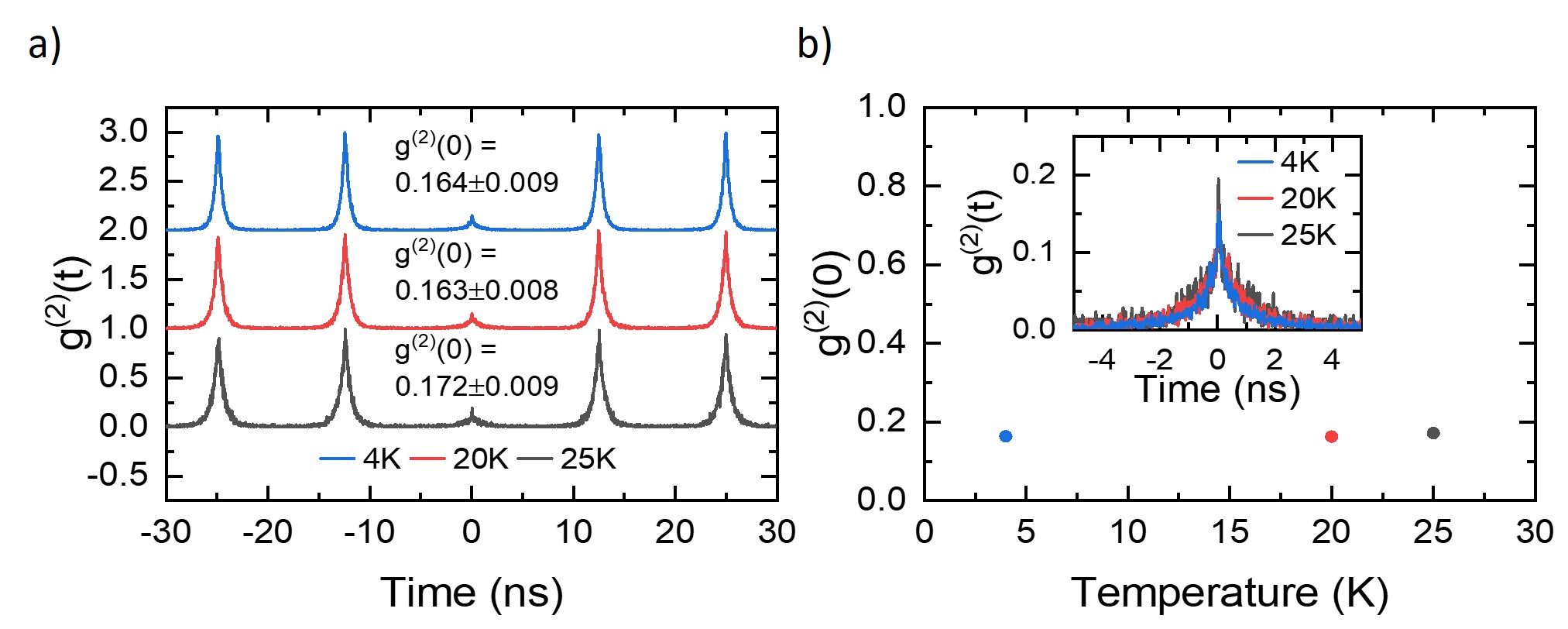}
\caption{a) Temperature dependent second-order correlation HBT measurements of the cavity coupled QD, measured using pulsed excitation through the QD phonon sideband. b) The g$^{(2)}$(0) values plotted against the sample temperature, showing the relative temperature insensitivity over 25K. Inset: magnification of the time-zero peak from (a), showing two components of the peak.}
\label{fig:HBT}
\end{figure}
To confirm that our device emits single photons, we performed Hanbury-Brown and Twiss (HBT) g$^{(2)}(t)$ autocorrelation measurements as a function of the sample temperature, again using pulsed quasi-resonant excitation via the phonon sideband. Figure (\ref{fig:HBT}a) shows the result of these measurements, with the data for all temperatures exhibiting clear antibunching through suppression of the peak at $t=0$. 
The inset of Figure (\ref{fig:HBT}b) shows the structure of this central time-zero peak at each temperature, normalised to the average of 5 peaks at non-zero delay. It can be seen that each of the time-zero peaks comprises two components. We attribute the sharp peak to the weak background also seen as the fast detector-response limited component in the lifetime measurements, whilst the temporally broader component is attributed to the QD transition. Fitting each peak of the HBT with a two-sided biexponential peak allows for extraction of the peak area. Comparing the area of the central time-zero peak to the neighbouring five peaks gives a g$^{(2)}$(0) of 0.164 $\pm$ 0.009 at 4.2 K. Raising the temperature to 25K sees minimal change (0.172 $\pm$ 0.009), showing that the HBT g$^{(2)}$(0) is insensitive to small temperature changes, within experimental accuracy. This is promising for applications where size, weight and power (SWAP) constraints may require cryocoolers with higher base temperatures, for example onboard satellites or in data centres.

\section*{Discussion}

The results of Figs. \ref{fig:fig_2} and \ref{fig:fig_3} demonstrate that our InP QD PhCC device exhibits a Purcell enhancement of $\sim 5$ with a corresponding QD radiative lifetime of 340 ps, despite the fit of Eq. \ref{eqn:Purcell} implying a spatial overlap between emitter and cavity of only $30 \%$. This value is comparable to the Purcell factors measured in bullseye resonator cavity structures where a $F_P$ of 3.0 $\pm$ 0.7 has been measured for a QD in an InAs/InGaAs/GaAs system\cite{Nawrath2023} and a $F_P$ of 6.7 $\pm$ 0.6 has been measured for InAs QDs embedded in an InAlGaAs membrane \cite{Kaupp2023}. An $F_P$ of 5 has been observed for a InAs QDs in an InP membrane in a line-defect PhCC, however the interpretation there is complicated by strong cavity mode emission\cite{Birowosuto2012}.

Setting $\epsilon = 1$ whilst keeping all other parameters fixed gives a theoretical maximum Purcell factor exceeding 30, corresponding to a QD radiative lifetime around 50 ps. We note that a cavity-QD device in the more technologically mature InGaAs platform has achieved a Purcell factor of 43 and a radiative lifetime of 23 ps in a similar low-Q PhCC nanostructure \cite{Liu2018}. Owing to their low mode volumes, this potential for large Purcell enhancements is a particular strength of PhCC structures, offering the opportunity to realise high repetition-rate single photon sources at telecommunications wavelengths that can enable high bit-rate quantum communications over fibre networks. Furthermore, we note that our results are obtained using LA phonon excitation, with the implementation of fast (quasi-)resonant driving previously proving essential to observe very short radiative lifetimes by elimination of relaxation effects from higher energy levels\cite{Liu2018}.

Meanwhile, the results of Fig. \ref{fig:HBT} demonstrate that our device exhibits the clear antibunching ($g^{(2)}(0) = 0.17$) that would be expected of a single photon source, and that this value is insensitive to temperature in the range 4 - 25 K. Lower g$^{(2)}(0)$ values have been obtained through LA phonon excitation of a metamorphic buffer C-band QDs \cite{Zeuner2021} in which a g$^{(2)}(0)$ of 0.038 $\pm$ 0.005 was observed. It has been shown that a $g^{(2)}(0)$ of 0.1 is tolerable for QKD purposes, achieved through p-shell excitation, using a QD SPS that emitted in the telecom o-band \cite{10.1063/5.0070966}. For longer range QKD communication an improved g$^{(2)}(0)$ would be required. The temporally narrow feature observed as a sharp spike in our $g^{(2)}$ data suggests a significant contribution from background emission possibly arising from QDs to lower energy. In the wafers used here the peak of the QD ensemble emission is centred at around 1700 nm. Shifting the emission closer to the C-band and thus reducing the background emission \cite{Sala2023} can improve $g^{(2)}(0)$ for DE InAs/InP QDs. This could open a pathway to a high purity, high repetition-rate SPS in the C-band for quantum communication and networking.

\section*{Conclusion}
We have demonstrated a short radiative lifetime of 0.343$\pm$ 0.001 ns for an MOVPE grown DE InAs QDs emitting at 1523 nm and coupled to an L3 PhCC. The short lifetime corresponds to a Purcell enhancement of 5 at 4.2 K. Through control of the sample temperature and laser excitation power, we show both control of the QD-cavity detuning and a preserved single photon emission purity at temperatures up to 25K. These findings suggest the viability of QD based, cryogen-free c-band single photon sources, supporting advancements in quantum communication technologies. As we noted earlier the QD transition measured using LA phonon sideband excitation was not observed under non-resonant, above band excitation at 850 nm. We believe that this is due to an unstable charge environment, reinforcing that electrical gating would be beneficial in future devices. Another benefit of electrical gating would be the reduction in QD transition shift due to DC Stark effect due to itinerant charges within the membrane. These changes coupled with a reduction in the sample background emission would improve this already promising SPS system.

\section*{Methods}

\subsection*{QD growth}
The samples investigated here are grown in a 3x2 close-coupled showerhead (CCS) Aixtron MOVPE reactor on InP(100) substrates, using H$_{2}$ as the carrier gas. The growth commences with a $\sim$300 nm InP buffer at 610$^o$C, followed by a sacrificial layer of Al$_{0.48}$In$_{0.52}$As, lattice-matched to InP. This is followed by a 310 nm InP membrane containing a single QD-layer in the middle.  The precursors used are trymethylindium (TMIn), trimethylgallium (TMGa), trymethylaluminium (TMAl) for group-III, and phosphine (PH$_{3}$) and arsine (AsH$_{3}$) for group-V. QDs are grown by DE as follows: 
Indium droplets are deposited on an In$_{0.719}$Ga$_{0.281}$As$_{0.608}$P$_{0.392}$ interlayer lattice-matched to InP\cite{Sala2023} at 400$^o$C with the with a TMIn flow of 1.4 $\mu$mol/min and exposed to an AsH$_{3}$ flow of 24 $\mu$mol/min for crystallization into QDs, while ramping the substrate temperature to 520$^o$C. The flows used for the InGaAsP layer are as follows: 32, 7.3, 310, and 7x10$^{3}$ $\mu$mol/min for TMIn, TMGa, AsH$_{3}$, and PH$_{3}$, respectively. After complete droplet crystallization, QDs are capped with a $\sim$20 nm low-temperature InP layer at the same temperature of 520$^o$C and followed by a further 135 nm InP grown at 610$^o$C to complete the membrane layer.  Thereafter, the samples are immediately cooled down and taken out of the growth chamber.
\subsection*{Sample fabrication}
After QD growth, a 300 nm thick SiN$_X$ hardmask was deposited on the wafer surface using silane-based plasma-enhanced chemical vapor deposition (Plasma-Therm 790 Series). The L3 PhCC nanophotonic design was then patterned on the wafer using e-beam lithography (Raith VOYAGER EBL system) combined with a thin film of AR-P 6200 e-beam resist. The pattern was transferred into the SiNx hardmask using a CHF$_3$-based reactive-ion etch (Plasma Technology RIE 80), and subsequently dry-etched into the InP membrane using a CH$_4$-H$_2$-Cl$_2$-based Inductively Coupled Plasma-RIE (Oxford Plasmalab System 100). After hardmask removal in buffered hydrofluoric acid the sacrificial layer was removed using a FeCl$_3$ wet etch.

\subsection*{QD and L3 cavity measurements}

The QD membrane sample was measured in a liquid helium bath cryostat with a base temperature of 4.2 K. For all measurements the excitation and collection locations were coincident on the sample.

Characterisation of the cavity modes was performed using above-band continuous wave (CW) excitation using a Thorlabs 852 nm diode laser. Measurements of $\tau_{bulk}$ were performed using pulsed above-band excitation at 850 nm using a Spectra Physics Tsunami ps laser with a pulse length of $\sim$2 ps. For cavity-enhanced lifetime and autocorrelation measurements with the LA-phonon sideband, we used an APE OPO-X fs with APE Pulse Slicer, pumped by a Spectra Physics Tsunami fs laser, which created a pulse $\sim$20 ps in duration. A cross-polarisation scheme and a WL Photonics 0.1 nm bandwidth, wavelength tunable filter was used to isolate the QD emission and filter out the excitation laser. Time-dependent measurements were performed using Single Quantum Superconducting Nanowire Single Photon Detectors (SNSPDs) and a Becker-Hickl SPC130 counting card.

For detuning-dependent measurements of the radiative lifetime a combination of sample heating and changing the applied excitation power was used to control the QD-cavity detuning. The sample temperature could be raised using a resistive heater and cernox temperature sensor (both Lake Shore) embedded in the sample mount. We attribute the laser power dependence of the transition wavelength to a DC stark shift induced by the photo-excitation of additional carriers in the bulk semiconductor at higher excitation powers. We also observed that the QD transition energy periodically shifted over time to longer wavelengths, allowing the effect of the cavity on the radiative lifetime to be mapped over a larger range of QD-cavity detunings. We hypothesise that the slow drift is also due to a DC Stark shift caused by charge instability near the QD. The absence of a diode structure with electrical control meant that the current sample had no mechanism to apply an external electric field to stabilise the charge environment.

\bibliography{sample}

\section*{Acknowledgements}
This work was funded by the EPSRC Quantum Communications Hub EP/T001011/1 and EPSRC Programme Grant EP/V026496/1.

\section*{Author contributions statement}
C.L.P., A.J.B., A.P.F. and A.T. performed the experiments. N.J.M., C.L.P. and A.P.F. designed and simulated the photonic crystal structure. M.G., R.D. and N.B. fabricated the photonic crystal structures. E.M.S. grew the QD wafer. L.W., J.H., M.S.S. and A.M.F. provided supervision and expertise. C.L.P., A.J.B., N.J.M., A.P.F., E.M.S. and R.D. wrote the manuscript with input from all authors.

\section*{Additional information}



\end{document}